%Paper: hep-ph/9206260
%From: PH520010@brownvm.brown.edu
%Date: Mon, 29 Jun 92 09:31:28 EDT

\input phyzzx

\def\3he{{$^3${\rm He}}}

\def\ie{{\it i.e.,\ }}

\def\slD{\raise.15ex\hbox{$/$}\kern-.53em\hbox{$D$}}
\def\slA{\raise.15ex\hbox{$/$}\kern-.53em\hbox{$A$}}
\def\dsl{\raise.15ex\hbox{$/$}\kern-.57em\hbox{$\Delta$}}
\def\slp{{\raise.15ex\hbox{$/$}\kern-.57em\hbox{$\partial$}}}
\def\nsl{\raise.15ex\hbox{$/$}\kern-.57em\hbox{$\nabla$}}
\def\sla{\raise.15ex\hbox{$/$}\kern-.57em\hbox{$\rightarrow$}}
\def\slla{\raise.15ex\hbox{$/$}\kern-.57em\hbox{$\lambda$}}
\def\slb{\raise.15ex\hbox{$/$}\kern-.57em\hbox{$b$}}
\def\slr{\raise.15ex\hbox{$/$}\kern-.57em\hbox{$r$}}
\def\lnp{\raise.15ex\hbox{$/$}\kern-.57em\hbox{$p$}}
\def\lnk{\raise.15ex\hbox{$/$}\kern-.57em\hbox{$k$}}
\def\lnK{\raise.15ex\hbox{$/$}\kern-.57em\hbox{$K$}}
\def\lnq{\raise.15ex\hbox{$/$}\kern-.57em\hbox{$q$}}
\def\nna{\raise.15ex\hbox{$/$}\kern-.57em\hbox{$a$}}

\def\a{\alpha}

\def\eps{{\epsilon}}

\def\la{\lambda}

\def\bbR{{I\kern-0.3em R}}

  %These three spacing deals will eventually
    % be fixed by Tony.

\def\fulltriangle{{{{{{{{{\pmb{\triangle}\kern-.65em\bullet}\kern-.4em
{\raise1.2ex\hbox{.}}}
\kern-.4em{\raise1.0ex\hbox{.}}}\kern-.2em{\raise1.0ex\hbox{.}}}
\kern-.4em{\raise.1ex\hbox{.}}}\kern-.4em{\raise.2ex\hbox{.}}}
\kern-.2em{\raise.35ex\hbox{.}}}\kern.1em{\raise.2ex\hbox{.}} }}
                                                                \

\def\hexagon{{\tenpoint
\langle\kern-.1em{\raise.2cm\hbox{$\overline{\hskip.7em\relax}$}}
\kern-.7em{\lower.3ex\hbox{$\underline{\hskip.7em\relax}$}}\kern-.075em
 \rangle}}

\def\pentagon{{\tenpoint
\raise.5ex\hbox{$\widehat{\qquad}$}\kern-1.8em{\backslash
\kern-.1em{\lower.3ex\hbox{$\underline{\kern.75em}$}}\kern-.05em/} }}

\def\pmb#1{\setbox0=\hbox{$#1$}%
\kern-.025em\copy0\kern-\wd0
\kern.05em\copy0\kern-\wd0
\kern-.025em\raise.0433em\box0 }

\def\q2{{Q^2}}
\def\gtwid{\raise.3ex\hbox{$>$\kern-.75em\lower1ex\hbox{$\sim$}}}
\def\ltwid{\raise.3ex\hbox{$<$\kern-.75em\lower1ex\hbox{$\sim$}}}
\def\12{{1\over2}}

\def\part{\partial}

\def\low#1{\lower.5ex\hbox{${}_#1$}}

\def\partt{\raise.15ex\hbox{$\widetilde$}{\kern-.37em\hbox{$\partial$}}}

\def\topppageno1{\global\footline={\hfil}\global\headline
={\ifnum\pageno<\firstpageno{\hfil}\else{\hss\twelverm --\ \folio
\ --\hss}\fi}}

\def\toppageno2{\global\footline={\hfil}\global\headline
={\ifnum\pageno<\firstpageno{\hfil}\else{\rightline{\hfill\hfill
\twelverm \ \folio
\ \hss}}\fi}}

\def\boxit#1{\vbox{\hrule\hbox{\vrule\kern3pt
  \vbox{\kern3pt#1\kern3pt}\kern3pt\vrule}\hrule}}

\rightline{BROWN-HET-865}

\title{Electroweak Baryogenesis with a Second Order Phase Transition
\foot{To be published in the proceedings of the March 1992 Texas-Yale Workshop
on Electroweak Baryogenesis, ed. by L. Krauss and S.-J. Rey (World
Scientific, Singapore, 1992)}}

\author{Robert H. Brandenberger$^{1)}$}
\vskip 1cm
\centerline{\rm and}
\author{Anne-Christine Davis$^{2)}$}
\vskip .5in
\centerline{\it {1) Physics Department, Brown University}}
\centerline{\it {Providence, RI 02912, USA}}
\vskip 1cm
\centerline{\it {2) Department of Applied Mathematics}}
\centerline{\it {and Theoretical Physics \& Kings College}}
\centerline{\it {University of Cambridge, Cambridge, CB39EW, U.K.}}

\abstract{
If stable electroweak strings are copiously produced
during the electroweak phase transition, they may contribute significantly
to the presently observed baryon to entropy ratio of the universe. This
analysis establishes the feasibility of implementing an electroweak
baryogenesis
scenario without a first order phase transition.
}

\endpage

\section{\bf Introduction}

The mechanisms suggested so far$^{1 - 4)}$ for electroweak baryogenesis all
rely on having a
first order phase transition. The resulting bubble walls were required in order
to obtain  a region of unsuppressed baryon number violation occurring out of
thermal equilibrium. Our work$^{5)}$ is based on the observation that
topological
defects forming in a second order phase transition may play a similar role to
the bubble walls. We propose a specific mechanism in which electroweak strings
are responsible for baryogenesis.

Electroweak strings$^{6)}$ are nontopological solitons which arise in the
standard electroweak theory (and extensions thereof). They are essentially
Nielsen-Olesen strings of $U(1)_Z$ embedded in the
 $SU(2)\times U(1)$
theory ($U(1)_Z$ is the Abelian subgroup which is broken during the electroweak
phase transition). For certain ranges of the parameters of the standard model,
electroweak strings are energetically stable$^{7)}$
(They are not topologically stable).

If, however, we are in a region of parameter space in which electroweak strings
are stable, a network of such strings will form during the electroweak phase
transition, even if it is second order. Inside the strings, anomalous baryon
number violating processes are unsuppressed. If the strings move, the out of
thermal equilibrium condition will be satisfied. Finally, the standard model
contains $CP$ violation. Hence all of Sakharov's criteria are satisfied. As we
shall demonstrate, it is in fact possible to generate a substantial $n_B/s$
using electroweak strings.

In order to obtain a sufficiently large baryon to entropy ratio, the standard
electroweak model must be extended by adding new terms in the Lagrangian which
contain explicit $CP$ violation. An often used prototype theory is the two
Higgs model.$^{2 - 4)}$

The construction of nontopological vortex solutions in theories which do
not satisfy the topological criterion for strings is not specific to the
minimal standard model. Thus, we expect electroweak strings to exist also in
extensions of the Weinberg-Salam model (This has recently been demonstrated
in the two Higgs model$^{8)}$). It is possible that these strings could
be stable even for experimentally allowed values for the model parameters. In
the following we shall assume that electroweak strings exist and are stable.

In models admitting stable electroweak strings, a network of such strings will
form during the electroweak phase transition. If we consider a theory with
Higgs potential
$$
V(\phi)=\la(\phi^+\phi-\eta^2/2)^2, \eqno\eq
$$
then the initial correlation length (mean separation of strings) will
be$^{9)}$
$$
\xi(t_G)\simeq\la^{-1}\eta^{-1}, \eqno\eq
$$
where $t_G$ is the time corresponding to the Ginsburg temperature of the phase
transition.

The initial network of electroweak strings will be quite different from that of
cosmic strings, the reason being that electroweak strings can end on local
monopole and antimonopole configurations. From thermodynamic
considerations$^{10)}$, we expect most of the strings to be short, \ie of
length $l\simeq\xi(t_G)$, since this maximizes the entropy of the network for
fixed energy.

After the phase transition, the vortices will contract along their axes and
decay after a time interval
$$
\Delta t_S\simeq{1\over v}(\la \eta)^{-1} \eqno\eq
$$
where $v$ is the velocity of contraction (expected to be $\simeq1$). In the
following, we shall demonstrate that the string contractions will produce a net
baryon symmetry. We are using units in which $c = \hbar = k_B = 1$.

\section{\bf The Baryogenesis Mechanism}

We shall consider an extension of the standard electroweak theory in which
there is additional $CP$ violation in the Higgs sector. An example is the two
Higgs model used in Refs.~2-4.
We assume that electroweak strings can be embedded in this model$^{8)}$,
and we choose
the values of the parameters in the Lagrangian for which these strings are
stable. Furthermore, the phase transition is taken to be second order.

A key issue is the formation probability of electroweak strings. In the
following, we make the rather optimistic assumption that both the mean length
and average separation of electroweak strings at $t_G$ will equal the
correlation length $\xi(t_G)$. For topological defects, this result follows
from the Kibble mechanism$^{9)}$. When applied to electroweak strings, the
Kibble mechanism implies that the vortex fields $\phi$ and $Z$ have the
correlation length $\xi(t_G)$. However, to form an electroweak string, the
other fields must be sufficiently small such that the configuration relaxes to
the exact electroweak string configuration. Obviously, the restriction this
imposes (and the consequent increase in the mean separation of electroweak
strings) is parameter dependent - the more stable the strings, the smaller
the increase in the mean separation.
Pieces of string are bounded by
monopole-antimonopole pairs. Energetic arguments tell us that the string will
shrink. We now argue that the moving string ends will have the same effects on
baryogenesis as the expanding bubble walls in Refs.~2\&3.

We can phrase our argument either in terms of the language of Ref. 2 or of
Ref. 3. The phase of the extra $CP$ violation is nonvanishing in the region in
which the Higgs fields $\phi$ are changing in magnitude, \ie at the edge of the
string. Since $|\phi|$ increases in magnitude, $CP$ violation has a definite
sign. Hence, in the language of Ref. 3, a chemical potential with definite sign
for baryon number is induced at the tips of the string (where $|\phi|$ is
increasing). This chemical potential induces a nonvanishing baryon number.

In the language of Ref. 2, the $CP$ violation with definite sign at the tips of
the string leads to preferential decay of local texture configurations with a
definite net change in Chern-Simons (\ie baryon) number.

Let us now estimate the magnitude of this effect. The rate of baryon number
violating events inside the string (in the unbroken phase) is
$$
\Gamma_B\sim\a_w^4T^4. \eqno\eq
$$
The volume in which $CP$ violation is effective changes at a rate ($g$ is the
gauge coupling constant)
$$
{dV\over dt}=g^2w^2V,\eqno\eq
$$
where $w\simeq\la^{-1/2}\eta^{-1}$ is the width of the string and $v$ is its
contraction velocity. The
factor $g$ comes from the observation that baryon number violating processes
are unsuppressed only if $|\phi|<g\eta$.$^{11)}$
The rate of baryon number generation per string is
$$
{dN_B\over dt}\sim w^2v \Gamma_B\eps\Delta t_c,\eqno\eq
$$
where $\eps$ is a dimensionless constant measuring the strength of $CP$
violation and
$$
\Delta t_c={gw\over v} {1\over \gamma(v)} \eqno\eq
$$
is the time a fixed point in space is in the transition region. Here,
$\gamma(v)$
is the usual relativistic $\gamma$ factor. Since there is one string per
correlation volume $\xi(t_G)^3$, the resulting rate of increase in the
baryon number density $n_B$ is
$$
{dn_B\over dt}\sim \la^{-3/2} \eta^{-3} g^3 \a_w^4 T^4 {1\over
\gamma(v)}
\eps \xi (t_G)^{-3}.\eqno\eq
$$
The net baryon number density is obtained by integrating (8) from $t_G$, the
time corresponding to the Ginsburg temperature, and $t_G+\Delta t_S$ (see (3)).
The result is
$$
n_B \sim {\la\over {v\gamma(v)}} g^3 \a_w^4 T^3_G\eps.\eqno\eq
$$

Our result (9) must be compared to the entropy density at $t_G$:
$$
s(t_G)= {\pi^2\over 45} g^* T^3_G,\eqno\eq
$$
where $g^*$ is the number of relativistic spin degrees of freedom. From (9)
and (10) we obtain
$$
{n_B\over s}\sim
{45\over\pi^2g^*}{\la\over {\gamma(v)v}}\eps g^3\a_w^4.\eqno\eq
$$
For $\la\sim v\sim1$ and $\eps\sim 1$, the ratio obtained is only slightly
smaller than the observational value.

In order for our mechanism to work, the core radius of the string ($\vert \phi
\vert < g \eta$) must be large enough to contain the nonperturbative
configurations which mediate baryon number violating processes. This leads to
the condition $\la < g^4$, i.e. small Higgs mass. In addition, the sphaleron
must be sufficiently heavy such that sphaleron transitions in the broken
symmetry phase are suppressed for $T = T_G$. For small values of $\la$, this
condition will automatically be satisfied. Finally, the model parameters must
be such that the phase transition is of second order. In the standard
electroweak theory, this condition is incompatible with $\la \ll g^4$. In any
extended electroweak theory, the consistency of the above conditions must be
satisfied in order for our baryogenesis mechanism to be effective.

\section{\bf Discussion}

We have presented a counterexample to the ``folk theorem" stating that
electroweak baryogenesis requires a first order electroweak phase transition.
We propose a mechanism in which finite length electroweak strings during their
contraction generate a nonvanishing net baryon number. The strings play
a similar role to the expanding bubble walls in a first order phase transition:
they provide out of equilibrium processes, and also a region where $CP$
violation occurs.

The mechanism presented here requires stable electroweak strings and an extra
source of $CP$ violation (which is present in the two Higgs models used in
Refs.~2-4). Based on the stability analysis of electroweak strings in the
standard model$^{7)}$, it is unlikely that these strings will be stable for
experimentally allowed values of the parameters in the Lagrangian.

\bigskip
\centerline{\bf Acknowledgements}

For interesting discussions we are grateful to
M. Einhorn,  R.~Holman and L.~McLerran. We also thank L. Krauss and S.-J. Rey
for organizing a stimulating and timely workshop.
 This
work was supported in part (at Brown) by DOE grant DE-AC02-76ER03130 Task A, by
an Alfred P. Sloan Fellowship (R.B.), and by an NSF-SERC Collaborative Research
Award NSF-INT-9022895 and SERC GR/G37149.
One of us (R.B.) acknowledges the
hospitality of the Institue for Theoretical Physics at the University of
California in Santa Barbara, where this work was completed with support from
NSF Grant No. PHY89-04035.

\bigskip
\centerline{\bf References}

\item{1.} M. Shaposhnikov, {\it JETP Lett.} {\bf 44}, 465 (1986);
M. Shaposhnikov, {\it Nucl. Phys.} {\bf B287} 757 (1987);
L. McLerran, {\it Phys. Rev. Lett.} {\bf 62}, 1075 (1989).

\item{2.} N. Turok and J. Zadrozny, {\it Phys. Rev. Lett.} {\bf 65} 2331
(1990); N. Turok and J. Zadrozny, {\it Nucl. Phys.} {\bf B358} 471 (1991);
L. McLerran, M. Shaposhnikov, N. Turok and M. Voloshin, {\it Phys. Lett.}
{\bf 256B}, 451 (1991).

\item{3.} A. Cohen, D. Kaplan and A. Nelson, {\it Phys. Lett.} {\bf 263B},
86 (1991).

\item{4.} A. Nelson, D. Kaplan and A. Cohen, {\it Nucl. Phys.} {\bf B373},
453 (1992).

\item{5.} R. Brandenberger and A.-C. Davis, {\it Electroweak Baryogenesis
with Electroweak Strings} Brown preprint BROWN-HET-862 (1992).

\item{6.} T. Vachaspati, {\it Phys. Rev. Lett.} {\bf 68}, 1977 (1992).

\item{7.} M. James, L. Perivolaropoulos and T. Vachaspati,
Tufts preprint (1992).

\item{8.} A.-C. Davis and U. Wiedemann, DAMTP preprint (1992).

\item{9.} T.W.B. Kibble, {\it J. Phys.} {\bf A9}, 1387 (1976).

\item{10.} D. Mitchell and N. Turok, {\it Phys. Rev. Lett.} {\bf 58}, 1577
(1987); M. Sakellariadou and A. Vilenkin, {\it Phys. Rev.} {\bf D42}, 349
(1990).

\item{11.} M. Dine, P. Huet, and R. Singleton, Santa Cruz preprint SCIPP
91/08 (1991).

\end